# Performance of Large Language Models in Technical MRI Question Answering: A Comparative Study

Alan B. McMillan

Department of Radiology
University of Wisconsin
Madison, Wisconsin
abmcmillan@wisc.edu

## Abstract

**Background**
Advances in artificial intelligence, particularly large language models (LLMs), have the potential to enhance technical expertise in magnetic resonance imaging (MRI), regardless of operator skill or geographic location.

**Methods**
We assessed the accuracy of several LLMs in answering 570 technical MRI questions derived from a standardized review book. The questions spanned nine MRI topics, including Basic Principles, Image Production, and Safety. Closed-source models (e.g., OpenAI's o1 Preview, GPT-4o, GPT-4 Turbo, and Claude 3.5 Haiku) and open-source models (e.g., Phi 3.5 Mini, Llama 3.1, smolLM2) were tested. Models were queried using standardized prompts via the LangChain framework, and responses were graded against correct answers using an automated scoring protocol. Accuracy, defined as the proportion of correct answers, was the primary outcome.

**Results**
The closed-source o1 Preview model achieved the highest accuracy (94%), exceeding the random-guess baseline (26.5%). GPT-4o and o1 Mini scored 88%, and GPT-4 Turbo and Claude 3.5 Haiku each scored 84%. Among open-source models, Phi 3.5 Mini performed well, achieving 78% accuracy, comparable to several closed-source models. Accuracy was highest in Basic Principles and Instrumentation categories but lower in Image Weighting and Contrast, History, and Artifacts and Corrections.

**Conclusions**
LLMs exhibit high accuracy in addressing technical MRI questions, suggesting their potential to standardize and enhance MRI practice. These models may improve image quality and consistency across varied clinical environments. Further studies are needed to refine LLMs for clinical use and integrate them into MRI workflows.

## 1. Introduction

Magnetic resonance imaging (MRI) is a sophisticated imaging modality that leverages powerful magnetic fields and radiofrequency pulses to generate detailed anatomical and functional images. Achieving optimal image quality with MRI requires substantial technical skill, as operator expertise directly influences the diagnostic accuracy and utility of the images produced. Variability in operator experience and training can lead to inconsistencies in image quality, which, in turn, may affect diagnostic reliability and patient outcomes [1], [2]. This variability in technical skill and knowledge is particularly evident in geographically isolated or resource-limited facilities, where access to experienced MRI technologists or radiologists may be limited. Thus, there is a need for tools that can provide consistent, expert-level guidance across diverse clinical settings, enhancing the uniformity of MRI practices worldwide.

The emergence of artificial intelligence (AI), particularly large language models (LLMs), offers a promising avenue to bridge this expertise gap in MRI. LLMs have demonstrated an unprecedented ability to interpret and generate human-like responses across a range of complex topics in radiology [3], [4], [5], suggesting their potential to provide real-time technical support in MRI. However, while these models show great promise, their utility in delivering precise, domain-specific guidance for MRI practices remains underexplored. The ability of LLMs to answer technical questions accurately could not only support less-experienced operators but also serve as a reliable resource for reinforcing best practices and enhancing overall image quality.

Variability in MRI quality has far-reaching implications for patient care, as poor imaging can lead to misdiagnoses, delayed treatment, and even unnecessary interventions. For instance, suboptimal image resolution or artifacts can obscure critical anatomical structures, potentially resulting in missed diagnoses of conditions such as small tumors, vascular abnormalities, or degenerative diseases [6], [7]. Studies have shown that inconsistencies in MRI protocol adherence, patient positioning, and artifact mitigation directly impact diagnostic accuracy and reliability [8]. Moreover, inadequate imaging quality can compromise longitudinal studies or treatment monitoring, leading to challenges in assessing therapeutic efficacy. Addressing these issues is critical to ensuring that MRI remains a reliable cornerstone of modern medical imaging.

Efforts to standardize MRI practices have traditionally relied on comprehensive training programs, certification processes, and adherence to established imaging protocols [9], [10], [11]. These approaches, while effective in structured settings, often fall short in resource-limited facilities where access to expert trainers, accreditation bodies, or continuous professional development opportunities may be scarce. The availability of highly trained technologists and graduate-level MRI physicists is generally not possible at every site [12], [13], [14]. Consequently, the variability in technical skill and imaging quality persists, highlighting the need for scalable, cost-effective tools that can provide real-time guidance to MRI operators, regardless of location or resource availability.

While large language models (LLMs) offer immense potential for improving MRI practices, their application in high-stakes medical settings is not without challenges. A major concern is the risk of generating hallucinated or inaccurate responses, which could mislead operators and compromise patient safety. Additionally, the closed-source nature of many advanced LLMs limits transparency, making it difficult to fully understand how these models derive their outputs or to verify the reliability of their recommendations. Ethical considerations, including data privacy and the potential for overreliance on AI tools, further complicate their deployment in clinical environments. These factors underscore the need for systematic evaluation of LLMs in domain-specific tasks to ensure their recommendations are accurate, interpretable, and actionable in diverse clinical contexts.

In this study, we sought to systematically evaluate the performance of multiple LLMs in responding to technical MRI questions. We hypothesized that advanced LLMs would exhibit high accuracy in addressing queries across a range of MRI-related topics and that larger, more sophisticated models would outperform smaller or less specialized models in their depth and accuracy of responses. By comparing the accuracy and relevance of responses from various LLMs, we aimed to assess their potential to serve as practical tools for supporting and standardizing technical expertise in MRI practice, especially in settings where direct access to expert guidance is limited. This research provides foundational insights into the potential of AI to support MRI operators in delivering consistent, high-quality imaging and could inform future developments in AI-powered technical support systems for radiology and beyond.

## 2. Materials and Methods

### 2.1 Study Design

This evaluative study aimed to systematically compare the performance of various large language models (LLMs) in answering technical questions specific to MRI. A diverse selection of models was chosen and are listed in Table 1 to represent a wide range of parameter sizes, training datasets, and developmental approaches, enabling analysis of their capabilities and limitations.

We evaluated a range of frontier, closed-source models. These included GPT, GPT Turbo, GPTo, GPTo-mini, o1, o1-mini, Gemini Flash, Gemini Pro, and Claude Haiku and Claude Sonnet. The GPT series models are widely recognized for their advanced natural language processing capabilities, with newer iterations noted for substantial improvements in contextual understanding and response generation. The Gemini models represent a line of advanced LLMs designed with a focus on performance and efficiency. The Claude models, developed with an emphasis on safety and reliability, provided a contrasting approach to AI model design and utility. Also included in the evaluation were small, open-source LLMs such as Gemma2 [15], Llama 3.1 [16], Mistral 7B [17], [18], Mistral NeMo [19], Nemotron Mini [20], Phi 3.5 [21], and smolLM2 [22]. These models were selected to assess the impact of parameter scale and open-source development on performance within a specialized domain like MRI.

| Vendor/Group | Model Name | Number of Parameters | Version | Source Availability |
|---|---|---|---|---|
| **Anthropic** | Claude 3.5 Haiku | Unknown | n/a* | Closed Source |
| **Anthropic** | Claude 3.5 Sonnet | Unknown | n/a* | Closed Source |
| **Google** | Gemini 1.5 Flash | Unknown | 002 | Closed Source |
| **Google** | Gemini 1.5 Flash 8B | 8 billion | n/a* | Closed Source |
| **Google** | Gemini 1.5 Pro | Unknown | 002 | Closed Source |
| **Google** | Gemma2 | 9 billion | n/a* | *Open Source* |
| **Hugging Face** | smolLM2 | 1.7 billion | n/a* | *Open Source* |
| **Meta** | Llama 3.1 | 8 billion | n/a* | *Open Source* |
| **Microsoft** | Phi 3.5 Mini | 3.8 billion | n/a* | *Open Source* |
| **Mistral AI** | Mistral 7B | 7 billion | 0.3 | *Open Source* |
| **Mistral AI & NVIDIA** | Mistral NeMo 12B | 12 billion | 2407 | *Open Source* |
| **NVIDIA** | Nemotron Mini 4B | 4 billion | n/a* | *Open Source* |
| **OpenAI** | GPT-3.5 Turbo | Unknown | 0125 | Closed Source |
| **OpenAI** | GPT-4 Turbo | Unknown | 2024-04-09 | Closed Source |
| **OpenAI** | GPT-4o | Unknown | 2024-08-06 | Closed Source |
| **OpenAI** | GPT-4o Mini | Unknown | 2024-07-18 | Closed Source |
| **OpenAI** | o1 Mini | Unknown | 2024-09-12 | Closed Source |
| **OpenAI** | o1 Preview | Unknown | preview-2024-09-12 | Closed Source |
| **Technology Innovation Institute** | Falcon 2 | 11 billion | n/a* | *Open Source* |

**Table 1.** List of language models evaluated for MRI question answering. *all models without version identifiers were accessed in November 2024.

### 2.2 Data Collection and Preparation

For this study, a comprehensive database of MRI-related questions was curated from *The MRI Study Guide for Technologists* [23], a well-regarded resource aimed at preparing technologists for professional registry exams. The study guide was selected for its content and structured presentation of technical concepts essential for MRI operation and troubleshooting. Questions and corresponding answers were systematically extracted from a digitized version of the book to facilitate consistent and accessible evaluation across models. To maintain clarity and avoid misinterpretations, questions that required visual reference to figures were excluded, as these could lead to ambiguity in a text-only prompt format. After this refinement, a total of 570 questions remained for analysis. These questions were then categorized into nine distinct MRI topics, representing a broad spectrum of foundational and advanced MRI concepts.

The final topic distribution was as follows: History (24 questions), covering the evolution of MRI technology; Basic Principles (64 questions), focused on foundational MRI physics; Image Weighting and Contrast (59 questions), addressing the mechanisms of image contrast; Image Production (115 questions), detailing the processes involved in generating MRI images; Pulse Sequences (41 questions), examining various sequences and their applications; Artifacts and Corrections (55 questions), exploring common imaging artifacts and their mitigation; Flow/Cardiac Imaging (82 questions), relevant to specialized imaging techniques for cardiovascular assessments; Instrumentation (56 questions), covering the technical components of MRI systems; and Safety (74 questions), emphasizing protocols and precautions critical to patient and operator safety.

### 2.3 LLM Evaluation Procedure

The LangChain framework [24] was employed to streamline and automate the process of querying large language models (LLMs) with MRI-related questions. This framework facilitated consistent formatting and sequential submission of each question to the different LLMs under evaluation. For open-source models listed in Table 1, local inference was performed using the Ollama platform [25] on a workstation with single 12 GB GPU card (NVIDIA Titan V). Closed-source models, such as Claude, Gemini, and GPT were accessed through respective cloud-based APIs, enabling consistent access to the latest model versions for each query. It remains unknown whether the content of the review book is included in the training data of any model tested. The policies of the closed-source models tested maintain a basic level of privacy, stipulating that inference data are not used for further training.

Each MRI-related question was presented to the models in a standardized text-only prompt format to minimize variability in interpretation. Model responses were captured systematically for subsequent grading and analysis. A uniform prompt structure was maintained across all models to control for prompt-related bias and to ensure that any observed differences in performance could be attributed to model characteristics rather than prompt variation.

### 2.4 Response Grading

An automated grading protocol was implemented to assess model accuracy by matching the choice selected by the language model to the correct answer extracted from the list of questions. The grading algorithm began by identifying the full text of the correct answer by matching the initial letter of the correct answer key to the corresponding answer choice. If the model's raw response was brief, consisting of fewer than three characters, the algorithm checked whether it began with the correct answer letter; if so, the response was considered correct. For longer responses, the algorithm determined whether the full text of the correct answer was present within the model's response, disregarding differences in case. In instances where a direct match was not found, fuzzy string matching using the Levenshtein distance was employed to compare the model's response with each of the answer choices [26], selecting the one with the highest similarity score as the given answer. This method permitted minor discrepancies in wording and focused on semantic alignment with the correct answer to determine accuracy. Each response was assigned a correctness label if it matched or was semantically equivalent to the reference answer.

### 2.5 Analysis

The primary metric of interest was the accuracy of each model, defined as the proportion of correctly answered questions across the total question set and within each MRI topic category. This accuracy measure allowed for a direct comparison of model performance on the specific technical content. A baseline "random guess" accuracy was calculated by averaging the expected success rate if answers were selected randomly, considering the average number of answer choices per question. This baseline provided context for interpreting each model's performance beyond random chance.

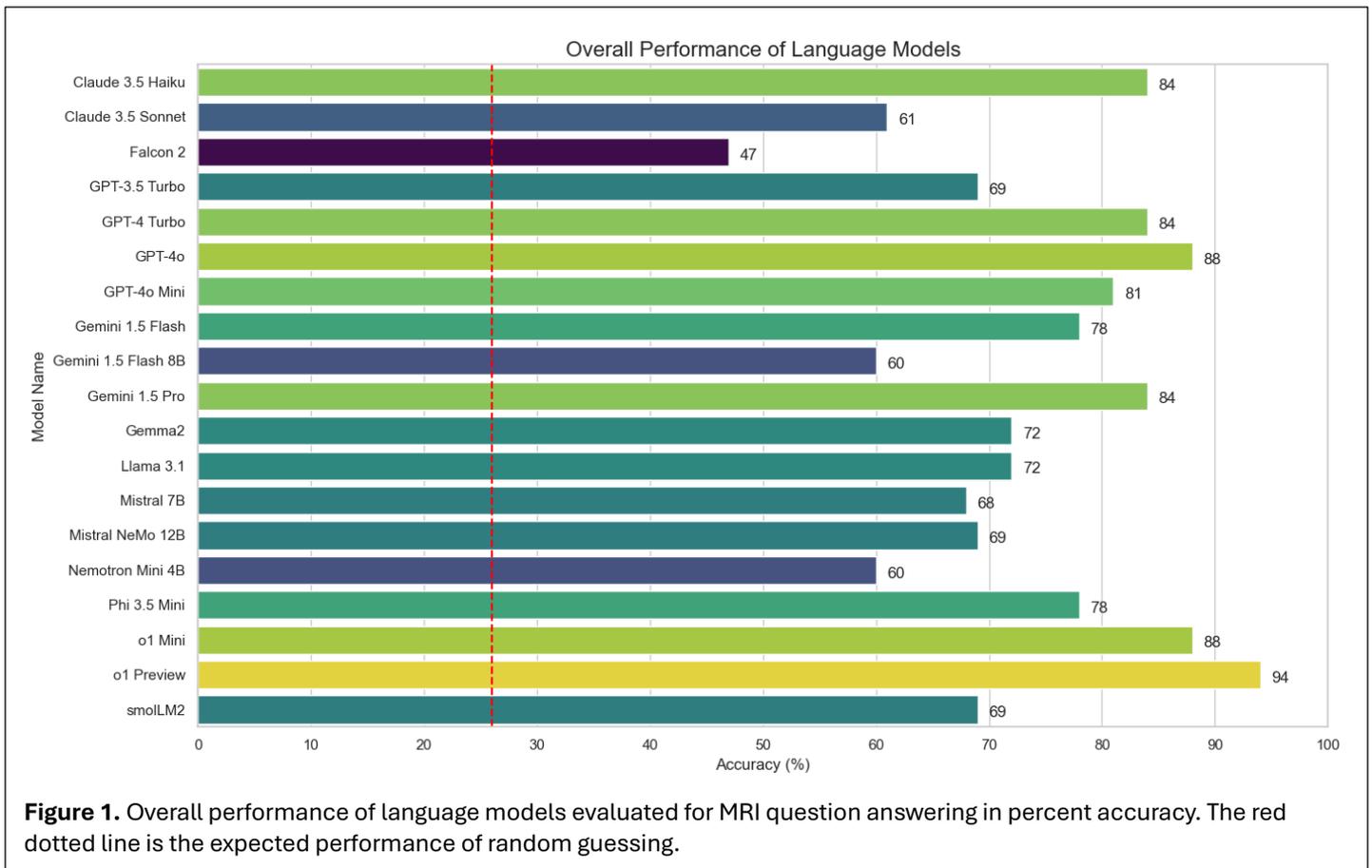

**Figure 1.** Overall performance of language models evaluated for MRI question answering in percent accuracy. The red dotted line is the expected performance of random guessing.

## 3. Results

In total, 570 questions were included in the analysis, covering nine distinct MRI topics. The average number of possible choices per question was approximately 3.78, resulting in an expected random guess accuracy of 26.5%. This baseline was calculated to provide context for evaluating each model's performance beyond chance alone. The overall percent accuracy for each model are shown in Figure 1, displaying the overall accuracy across the question set. OpenAI's o1 Preview model achieved the highest overall accuracy at 94%, followed closely by GPT-4o and o1 Mini, both at 88%. GPT-4 Turbo and Anthropic's Claude 3.5 Haiku both scored 84%. These closed-source models, known for their advanced architectures and extensive pre-training, outperformed other models by a substantial margin. Among other models, Google's Gemini 1.5 Pro also achieved a high accuracy of 84%, while Gemini 1.5 Flash and Microsoft's Phi 3.5 Mini scored 78%, which was the best performing open-source model tested. Other open-source models such as Meta's Llama 3.1 and Hugging Face's smolLM2 achieved moderate accuracies of 72% and 69%, respectively. Technology Innovation Institute's Falcon 2 had a lower accuracy of 47%.

A heatmap showing the performance across the 9 categories is shown in Figure 2. Analyzing the performance across individual MRI categories, OpenAI's o1 Preview model consistently achieved the highest accuracy in each area. In the History category, o1 Preview tied for the top score of 92% with OpenAI's o1 Mini and Google's Gemini 1.5 Pro. For Basic Principles, o1 Preview achieved the highest accuracy at 97%, outperforming other models like GPT-4o and Claude 3.5 Haiku, which scored 94%. In Image Weighting and Contrast, both o1 Preview and o1 Mini led with a top score of 81%. In the Image Production category, o1 Preview stood out with a leading accuracy of 96%, surpassing GPT-4o's 89% and Claude 3.5 Haiku's 83%. The model also attained the highest accuracies in Pulse Sequences (95%),

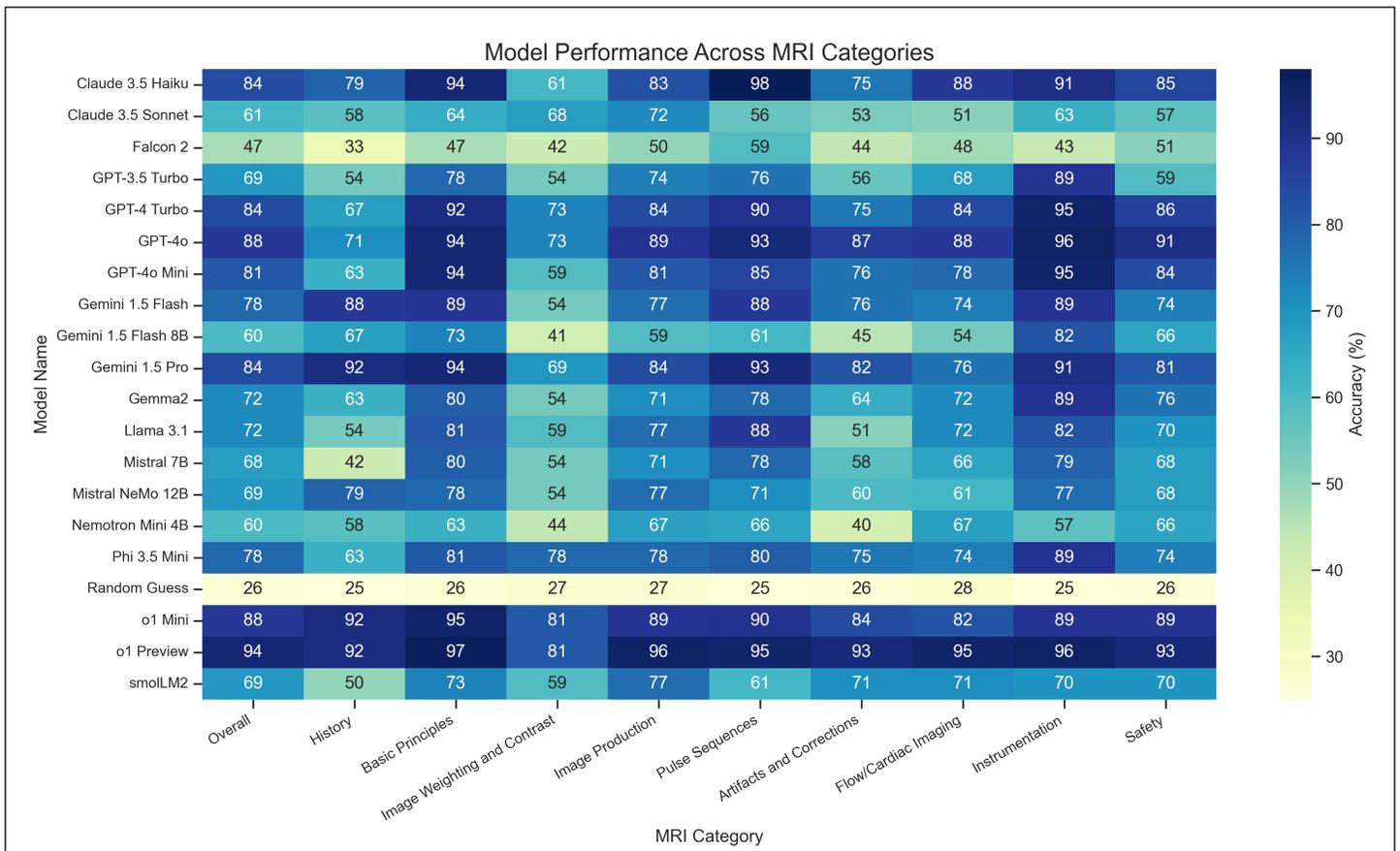

**Figure 2.** Performance of language models evaluated for MRI question answering in percent accuracy. Scores include Overall and category-wise scores. All models performed better than a random guess. o1 Preview performed the best compared to all other models tested.

Artifacts and Corrections (93%), and Flow/Cardiac Imaging (95%). In the Instrumentation category, o1 Preview tied with OpenAI's GPT-4o, both scoring 96%, indicating exceptional understanding of MRI hardware components. Finally, in the Safety category, o1 Preview achieved the top score of 93%, slightly ahead of GPT-4o's 91% and Claude 3.5 Haiku's 85%.

Analyzing the models' performance across different MRI categories reveals that, overall, they performed worst in the Image Weighting and Contrast, History, and Artifacts and Corrections categories. The Image Weighting and Contrast category had the lowest average accuracy among the models, indicating challenges in comprehending the nuanced principles that determine MRI image properties and contrast mechanisms. Similarly, the History category showed lower performance, suggesting that models may have limited knowledge of the historical developments and foundational milestones in MRI technology. The Artifacts and Corrections category also posed difficulties, reflecting the complexity involved in identifying and mitigating imaging artifacts within MRI diagnostics.

## 4. Discussion

This study provides a comparative analysis of various large language models (LLMs) in their capacity to accurately answer technical questions related to MRI. Among the models evaluated, OpenAI's o1 Preview demonstrated the highest accuracy, achieving 94%, which markedly surpassed other models and significantly exceeded the random guess baseline of 26.5%. The superior performance of o1 Preview, alongside other closed-source models such as GPT-4o and Claude 3.5 Haiku, underscores the importance of advanced architectures and extensive pre-training in achieving higher accuracy. These findings suggest that models with sophisticated training and larger parameter counts are better equipped to manage the complexities inherent in technical MRI knowledge. The high accuracy achieved by OpenAI's o1 Preview highlights the advantages of advanced models with extensive pre-training on varied datasets, as such models appear better able to capture and interpret the nuanced technical information required in MRI contexts. In contrast, open-source models, which have drastically smaller parameter counts, had generally lower performance. However, the open-source Phi3.5 model performed equivalently to the closed source Gemini 1.5 Flash (78% overall accuracy), and exceeded Gemini 1.5 Flash 8B. Furthermore, the smallest model tested, smolLM2, matched the overall performance of GPT-3.5 (69% overall accuracy), demonstrating that small efficient models can be as capable as much larger models.

The strong performance of language models overall in categories such as Basic Principles and Instrumentation reflects an ability to process and relay foundational and factual information accurately. However, the models faced challenges in more complex areas like Image Weighting and Contrast, and Artifacts and Corrections, indicating potential gaps in the depth of their contextual and domain-specific understanding.

The demonstrated ability of OpenAI's o1 Preview to accurately respond to technical MRI questions highlights its potential for applications in clinical MRI settings. Such LLMs could serve as on-demand reference tools for MRI technologists, providing real-time guidance on technical issues, assisting with protocol adjustments, and offering solutions to mitigate common imaging artifacts. By delivering rapid access to accurate technical information, advanced LLMs have the potential to improve consistency and quality in MRI examinations, which could be particularly beneficial in settings with limited access to specialized expertise. Integrating these models into clinical practice could also streamline the learning process for new technologists and enhance continuous education for seasoned operators, supporting standardization of MRI practices across institutions and ultimately contributing to improved patient care outcomes.

This study has several limitations that warrant careful consideration. Relying solely on questions from a single review book may limit the diversity of MRI concepts assessed, potentially affecting the generalizability of the findings to broader or more nuanced aspects of MRI practice. Additionally, the closed-source nature of top-performing models, such as OpenAI's o1 Preview and GPT-4o, restricts transparency; proprietary details about their training datasets and architectures hinder a deeper understanding of the factors influencing their performance. Furthermore, none of the models were fine-tuned on MRI-specific data, which could have limited the accuracy of the open-source models; fine-tuning might improve their capabilities, helping to narrow the gap between them and more advanced models. Moreover, the evaluation was based exclusively on multiple-choice questions, which may not fully capture the models' proficiency in more complex or open-ended MRI-related tasks.

The findings of this study underscore the potential for advanced LLMs to serve as valuable resources for MRI practice by accurately addressing technical queries. To further refine and expand the role of LLMs in this domain, future research should pursue domain-specific fine-tuning, training models on specialized MRI datasets to enhance their understanding of complex concepts unique to this field. Human evaluation of model responses by MRI experts would provide a deeper assessment of their applicability, aiding in the identification of strengths and weaknesses from a practical standpoint. Additionally, investigating pathways for integrating LLMs into clinical workflows would clarify how these models could effectively support technologists in real-time, fostering improved decision-making and technical troubleshooting. Lastly, ethical and practical considerations, such as issues surrounding accuracy, accountability, and the impact of LLM reliance on technologist training and employment, warrant careful exploration to ensure that AI integration supports both patient care and the professional development of MRI personnel.

## 5. Conclusion

Advanced large language models like GPT-4 exhibit high accuracy in answering technical MRI questions,

indicating their potential to enhance technical expertise in MRI practice. By providing real-time support and standardized knowledge, LLMs could improve image quality, efficiency, and patient outcomes. Further research and development are needed to optimize these models for clinical application.